\documentclass[twocolumn,superscriptaddress,showpacs,preprintnumbers,amsmath,amssymb]{revtex4}
\usepackage{graphicx}% Include figure files
\usepackage{dcolumn}% Align table columns on decimal point
\usepackage{bm}% bold math
\usepackage{longtable}
\def\bea {\begin{eqnarray}}
\def\eea {\end{eqnarray}}

\def\be {\begin{equation}}
\def\ee {\end{equation}}

\begin{document}
\title{Magnetic dipole excitations of $^{50}$Cr}
\author{H.~Pai}
\email{hari.vecc@gmail.com,hpai@ikp.tu-darmstadt.de}
\affiliation{Institut f\"ur Kernphysik, Technische Universit\"at Darmstadt, D-64289 Darmstadt, Germany}
\author{T.~Beck}
\affiliation{Institut f\"ur Kernphysik, Technische Universit\"at Darmstadt, D-64289 Darmstadt, Germany}
\author{J.~Beller}
\affiliation{Institut f\"ur Kernphysik, Technische Universit\"at Darmstadt, D-64289 Darmstadt, Germany}
\author{R.~Beyer}
\affiliation{Institut f\"ur Strahlenphysik, Helmholtz-Zentrum Dresden-Rossendorf, D-01328 Dresden, Germany}
\author{M.~Bhike}
\affiliation{Department of Physics, Duke University, Durham, North Carolina 27708, USA}
\affiliation{Triangle Universities Nuclear Laboratory, Durham, North Carolina 27708, USA}
\author{V.~Derya}
\affiliation{Institut f\"ur Kernphysik, Universit\"at zu K\"oln, D-50937 K\"oln, Germany}
\author{U.~Gayer}
\affiliation{Institut f\"ur Kernphysik, Technische Universit\"at Darmstadt, D-64289 Darmstadt, Germany}
\author{J.~Isaak}
\affiliation{ExtreMe Matter Institute EMMI and Research Division, GSI Helmholtzzentrum f\"ur Schwerionenforschung GmbH, D-64291 Darmstadt, Germany}
\affiliation{Frankfurt Institute for Advanced Studies FIAS, D-60438 Frankfurt am Main, Germany}
\author{Krishichayan}
\affiliation{Department of Physics, Duke University, Durham, North Carolina 27708, USA}
\affiliation{Triangle Universities Nuclear Laboratory, Durham, North Carolina 27708, USA}
\author{J. Kvasil}
\affiliation{Institute of Particle and Nuclear Physics, Charles University, CZ-18000, Prague 8, Czech Republic}
\author{B.~L$\rm \ddot{o}$her}
\affiliation{Institut f\"ur Kernphysik, Technische Universit\"at Darmstadt, D-64289 Darmstadt, Germany}
\affiliation{ExtreMe Matter Institute EMMI and Research Division, GSI Helmholtzzentrum f\"ur Schwerionenforschung GmbH, D-64291 Darmstadt, Germany}
\author{V.~O.~Nesterenko}
\affiliation{Laboratory of Theoretical Physics, Joint Institute for Nuclear Research, Dubna, Moscow region, 141980, Russia}
\author{N.~Pietralla}
\affiliation{Institut f\"ur Kernphysik, Technische Universit\"at Darmstadt, D-64289 Darmstadt, Germany}
\author{G.~Mart\'inez-Pinedo}
\affiliation{Institut f\"ur Kernphysik, Technische Universit\"at
  Darmstadt, D-64289 Darmstadt, Germany}
\affiliation{GSI Helmholtzzentrum f\"ur Schwerionenforschung,
  Planckstra{\ss}e~1, D-64291 Darmstadt, Germany}
\author{L.~Mertes}
\affiliation{Institut f\"ur Kernphysik, Technische Universit\"at Darmstadt, D-64289 Darmstadt, Germany}
\author{V.~Yu.~Ponomarev}
\affiliation{Institut f\"ur Kernphysik, Technische Universit\"at Darmstadt, D-64289 Darmstadt, Germany}
\author{P.-G.~Reinhard}
\affiliation{ Institut f\"ur Theoretische Physik II,
Universit\"at Erlangen, D-91058, Erlangen, Germany}
\author{A. Repko}
\affiliation{Institute of Particle and Nuclear Physics, Charles University, CZ-18000, Prague 8, Czech Republic}
\author{P.~C.~Ries}
\affiliation{Institut f\"ur Kernphysik, Technische Universit\"at Darmstadt, D-64289 Darmstadt, Germany}
\author{C.~Romig}
\affiliation{Institut f\"ur Kernphysik, Technische Universit\"at Darmstadt, D-64289 Darmstadt, Germany}
\author{D.~Savran}
\affiliation{ExtreMe Matter Institute EMMI and Research Division, GSI Helmholtzzentrum f\"ur Schwerionenforschung GmbH, D-64291 Darmstadt, Germany}
\affiliation{Frankfurt Institute for Advanced Studies FIAS, D-60438 Frankfurt am Main, Germany}
%%\author{M.~Scheck$^{1,2,3}$}
%%\altaffiliation{Present Address: School of Engineering, University of the West of Scotland, PA1 2BE Paisley, UK;~
%%SUPA, Scottish Universities Physics Alliance, Glasgow G12 8QQ, UK}
%%\author{L. Schnorrenberger$^{1}$}
%\author{S.~Volz$^{1}$}
\author{R.~Schwengner}
\affiliation{Institut f\"ur Strahlenphysik, Helmholtz-Zentrum Dresden-Rossendorf, D-01328 Dresden, Germany}
\author{W.~Tornow}
\affiliation{Department of Physics, Duke University, Durham, North Carolina 27708, USA}
\affiliation{Triangle Universities Nuclear Laboratory, Durham, North Carolina 27708, USA}
\author{V.~Werner}
\affiliation{Institut f\"ur Kernphysik, Technische Universit\"at Darmstadt, D-64289 Darmstadt, Germany}
\author{J.~Wilhelmy}
\affiliation{Institut f\"ur Kernphysik, Universit\"at zu K\"oln, D-50937 K\"oln, Germany}
\author{A.~Zilges}
\affiliation{Institut f\"ur Kernphysik, Universit\"at zu K\"oln, D-50937 K\"oln, Germany}
\author{M.~Zweidinger}
\affiliation{Institut f\"ur Kernphysik, Technische Universit\"at Darmstadt, D-64289 Darmstadt, Germany}
%\affiliation{%
%$^1$Institut f\"ur Kernphysik, Technische Universit\"at Darmstadt, D-64289 Darmstadt, Germany\\
%%$^2$School of Engineering, University of the West of Scotland, PA1 2BE Paisley, UK\\
%%$^3$SUPA, Scottish Universities Physics Alliance, Glasgow G12 8QQ, UK}%
%$^2$Department of Physics, Duke University, Durham, North Carolina 27708, USA\\
%$^3$Triangle Universities Nuclear Laboratory, Durham, North Carolina 27708, USA\\
%$^4$Institut f\"ur Kernphysik, Universit\"at zu K\"oln, D-50937 K\"oln, Germany\\
%$^5$ExtreMe Matter Institute EMMI and Research Division, GSI Helmholtzzentrum f\"ur Schwerionenforschung GmbH, D-64291 Darmstadt, Germany\\
%$^6$Frankfurt Institute for Advanced Studies FIAS, D-60438 Frankfurt am Main, Germany\\
%$^7$Laboratory of Theoretical Physics, Joint Institute for Nuclear Research, Dubna, Moscow region, 141980, Russia\\
%$^8$Institut f\"ur Strahlenphysik, Helmholtz-Zentrum Dresden-Rossendorf, 01328 Dresden, Germany}%
%\author{Haridas Pai}
%% \homepage{}
%\affiliation{TU Darmstadt}
%%\\
 %%forced% with \\
%%}%
\date{\today}% It is always \today, today,
             %  but any date may be explicitly specified

\begin{abstract}
The low-lying $M1$-strength of the open-shell nucleus $^{50}$Cr has been studied with the method of nuclear resonance fluorescence up to 9.7~MeV, using
bremsstrahlung at the superconducting Darmstadt linear electron accelerator S-DALINAC and Compton backscattered photons at the High Intensity $\gamma$-ray
Source (HI$\gamma$S) facility between 6 and 9.7 MeV of the initial photon energy. Fifteen $1^{+}$ states have been observed between 3.6 and 9.7 MeV.
Following our analysis, the lowest $1^{+}$ state at 3.6 MeV can be considered as an isovector orbital mode with some spin admixture. The obtained results
generally match the estimations and trends typical for the scissors-like mode. Detailed calculations
 within the Skyrme Quasiparticle Random-Phase-Approximation method and the Large-Scale Shell Model
 justify our conclusions. The calculated distributions
 of the orbital current for the lowest $1^{+}$-state suggest the schematic view
 of Lipparini and Stringari (isovector rotation-like oscillations inside the rigid surface)
  rather than the scissors-like picture
 of Lo Iudice and Palumbo. The spin M1 resonance is shown to be mainly generated
by spin-flip transitions between the orbitals of the $fp$-shell.
\end{abstract}
\pacs{21.10.Re; 23.20.Lv; 25.20.Dc; 21.60.Jz; 27.50.+e }

\maketitle
\section{Introduction}

During the last decades, investigation of strong magnetic dipole ($M1$)
transitions in medium-mass nuclei was of a keen interest
\cite{ar,Har01,dz,npr,rev,Hei10}. First of all, this was caused by a diversity of
various physical effects related to these excitations: quenching of
spin strength and impact of non-nucleonic degrees of freedom,
relation to Gamow-Teller transitions and relevant astrophysical
problems, isospin degrees of freedom of valence-shell excitations,
peculiarities of the orbital scissors-like mode, cross-shell and
l-forbidden transitions, impact of complex configurations, etc.
Furthermore, these nuclei are accessible to modern large-scale shell model calculations
\cite{lssm_Talmi,lssm_Gaurier} that have the capability to
describe the effects mentioned above.
Theoretical efforts and advanced experiments have finally led to a significant
progress in understanding the features
of $M1$ excitations in this mass region \cite{Hei10}.

Open-shell $fp$-shell nuclei are particularly important in this activity \cite{fe,cu,mn}.
Description  of these nuclei is rather complicated because here deformation and pairing effects
come to play. At the same time, these effects make the physics of $fp$-shell nuclei much richer.
In particular, deformation results in appearance of the low-energy orbital scissors-like $M1$ mode
($1^{+}_{sc}$) \cite{Iudice,ic,db,NLI97}.

So far the experimental data on the $M1$ $1^{+}_{sc}$ strength in the $fp$-region were
limited to a few axially deformed nuclei: $^{46,48}$Ti (Z=22)\cite{ar} and $^{56}$Fe (Z=26) \cite{fe}.
It was found that the $1^{+}_{sc}$ in these nuclei is mainly represented by one low-energy (in the region 3.4-4.4 MeV)
state with $K^{\pi}=1^+$. As expected \cite{Iudice,ic,db,NLI97}, this state is characterized
by a strong $M1$ transition to the ground state. The theoretical analysis~\cite{ar,li,fe} indicates
that the $1^{+}_{sc}$ in this mass region is of a mixed
(orbital + spin) character, although with a
dominant orbital contribution. Shell-model calculations \cite{fe} demonstrate a considerable
dependence of the description on a subtle balance of different effects such as
equilibrium deformation, pairing, interaction parameters, etc.

For a better understanding of the $1^{+}_{sc}$ features in $fp$-shell nuclei,
we certainly need experimental data for deformed Cr (Z=24) isotopes placed
between the Ti and Fe chains. $J^{\pi}=1^+$ states of $^{50}$Cr are known from
previous ($p$,$p^\prime$) experiments~\cite{ppr} and for three of them
$\gamma$-ray transitions have been measured in nuclear resonance fluorescence
(NRF)~\cite{en,rm}. $B(M1)$ values are known only for two $1^{+}$ states of $^{50}$Cr. A comprehensive $M1$ strength distribution is missing. It
is the purpose of the present study to provide data for a more comprehensive $M1$
strength distribution for the deformed nucleus $^{50}$Cr.

This nucleus was studied in photon scattering experiments using NRF~\cite{uk,np,fr}.
The experiment was performed at the superconducting
Darmstadt electron linear accelerator S-DALINAC~\cite{sdl} using unpolarized
bremsstrahlung with a continuous spectral distribution from
the Darmstadt High Intensity Photon Setup (DHIPS)~\cite{ks}. Parity quantum numbers of spin-1 states were
determined using the High Intensity $\gamma$-Ray Source~\cite{higs1} operated by the
Triangle Universities Nuclear Laboratory and the Duke Free Electron Laser Laboratory (DFELL)
at Duke University in Durham, NC, USA.

In the present experiment, the spectrum of $^{50}$Cr up to 9.7 MeV and
the corresponding M1 strength were determined. The
M1 strength for the lowest $K^{\pi}=1^+$-state with excitation energy of 3628.2
keV was obtained and the $B(M1)$ values for 12 strongly excited $1^+$ states were measured for the first time.

Two theoretical models were used for further interpretation:
Skyrme Quasiparticle Random-Phase-Approximation (QRPA)~\cite{Ring,Repko} and Large-Scale Shell Model
(LSSM)~\cite{lssm_Gaurier}. The self-consistent QRPA and LSSM rather well
reproduce the experimental axial quadrupole deformation
$\beta_{\rm{exp}}$=0.2897(44) \cite{bnl} of $^{50}$Cr and provided
qualitatively close results. The calculations agree upon the feature that the 3628.2-keV state
has predominantely orbital character with a minor spin admixture. This means
that the present experiment provides observation of the $1^{+}_{sc}$ in Cr
isotopes. The new data essentially supplement the $1^{+}_{sc}$ systematics in
$fp$-shell nuclei and, as shown below, lead to a better understanding
of $1^{+}_{sc}$ features in this mass region.
Besides our analysis of the orbital flow implies a closer correspondence
to the schematic view
of Lipparini and Stringari~\cite{lip_str_83} corresponding to isovector
rotation-like oscillations of the nucleons inside the rigid nuclear surface
than to the scissors-like picture of Lo Iudice and Palumbo \cite{Iudice}.

The paper is organized as follows. In Sec. II, the experimental method and data analysis are outlined.
In Sec. III, the theoretical analysis of the results is given. Sec. IV provides a summary.

%%%%%%%%%%%%%%%%%%%%%%%%%%%%%%%%%%%%%%%%%%%%%%%%%%%%%%%%%%%%
\section{Experimental Method and Data Analysis}

In NRF measurements, the excitation mechanism is purely electromagnetic.
Therefore, intrinsic properties like spin, parity, and transition
probabilities can be determined
from the measured quantities (angular distribution, polarization asymmetries,
$\gamma$-ray energy, and intensity~\cite{uk,np,fr}) in a model-independent way.
For details of the NRF method, basic relations between the detected number
of events and energy-integrated cross-sections,
transition widths as well as $M1$ transition strengths, we refer
the reader to the reviews by Metzger~\cite{fr} and Kneissl
and coworkers~\cite{uk,np}. Spin quantum numbers, cross-sections $I_{i,0}$ (energy-integrated scattering cross sections for exciting a state $i$ and
deexciting this state to the ground state 0), ground-state
transition widths ${\Gamma_0}$, and transition strengths $B(M1)$ were measured
at the DHIPS setup.

A series of NRF experiments was performed at TU Darmstadt and at Duke University.
The photon-scattering experiment with unpolarized bremsstrahlung from the Darmstadt
linear electron accelerator S-DALINAC~\cite{sdl} was
performed at the Darmstadt High Intensity Photon Setup
(DHIPS)~\cite{ks}. Two measurements on $^{50}$Cr were performed with
bremsstrahlung at endpoint energies of 7.5(1)~MeV and 9.7(1)~MeV, respectively.
The measurement with 7.5(1) MeV bremsstrahlung has
been carried out to identify transitions via intermediate states.
Bremsstrahlung has been produced by completely stopping the intense
electron beam from the S-DALINAC's injector in a thick copper radiator target.
The generated photon beam passes a massive copper collimator resulting in a beam spot
with a size of about 2.5 {cm} diameter at the NRF target position.
Target nuclei are
excited by the resonant absorption of photons and subsequently decay either directly
or via intermediate states to the ground state.

Scattered $\gamma$-rays were detected by three HPGe detectors with 100\% efficiency relative to
a standard $3\verb+"+ \times 3\verb+"+$ NaI detector at a $\gamma$-ray energy of 1.3 {MeV}.
They are placed at polar angles of $90^{\circ}$ and $130^{\circ}$
with respect to the incident beam.
The detectors were surrounded by lead and  BGO Compton suppression shields.
The first measurement was performed with 2.0 g of an isotopically enriched
$^{50}$Cr target (96.416\% enriched) using
an end-point energy of $E_{0}$~=~9.7(1)~MeV and
average electron beam currents of about 20 $\mu$A. In this measurement data were
taken for about 131 hours.
For energy and photon-flux calibrations 399.3 mg of enriched
$^{11}$B (99.52\% enriched) were used that were irradiated simultaneously.
The efficiency of the two HPGe detectors was determined with a $^{56}$Co source up to about
3500 keV energy. For higher photon energies the gamma-ray detection
efficiencies were extracted from a simulation using the Geant4 toolkit~\cite{g4}.
The second measurement was performed with the same target mass using an
end-point energy at $E_{0}$~=~7.5(1)~MeV and
average electron beam currents of about 33 $\mu$A. In this measurement, data were
taken for about 105 hours.
%%%%%%%%%%%%%%%%%%%%%%
% Figure 1
%%%%%%%%%%%%%%%%%%%%%
\begin{figure}[htb]
  \begin{center}
    \includegraphics*[angle=0,width=\linewidth]{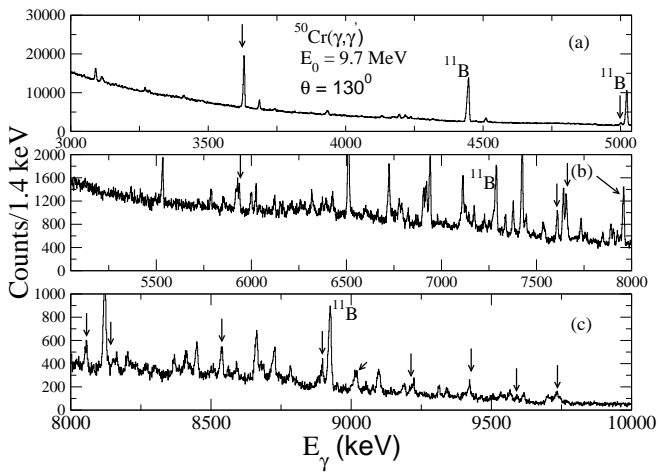}
    \caption{Photon scattering spectra of the
      $^{50}$Cr($\gamma,\gamma^\prime$) reaction from DHIPS between
      3000 keV and 10000 keV for the $130^{\circ}$ detector, measured
      at $E_{0}$~=~9.7~MeV. Ground-state $M1$ transitions of $^{50}$Cr
      are indicated by arrows. \label{spectrum}}
  \end{center}
\end{figure}

Photon scattering spectra of the $^{50}$Cr($\gamma,\gamma^{\prime}$)
reaction from DHIPS between 3000 and 10000 keV for the detector at  $130^{\circ}$,
measured at $E_{0}$~=~9.7(1)~MeV are shown in Fig.~\ref{spectrum}.
We have observed 33 excited states of $^{50}$Cr with spin quantum numbers $J=1$.
In this work, we focus on $J^{\pi}$=$1^+$ states.
$M1$ transitions to the ground state of $^{50}$Cr are indicated in Fig.~\ref{spectrum} by arrows.
In the following, transitions corresponding to direct decays to the ground state are
called elastic transitions and those decaying via intermediate states will be referred to as inelastic transitions.
%%%%%%%%%%%%%%%%%%%%%%
% Figure 2
%%%%%%%%%%%%%%%%%%%%%
\begin{figure}[htb]
  \begin{center}
    \includegraphics*[width=\linewidth]{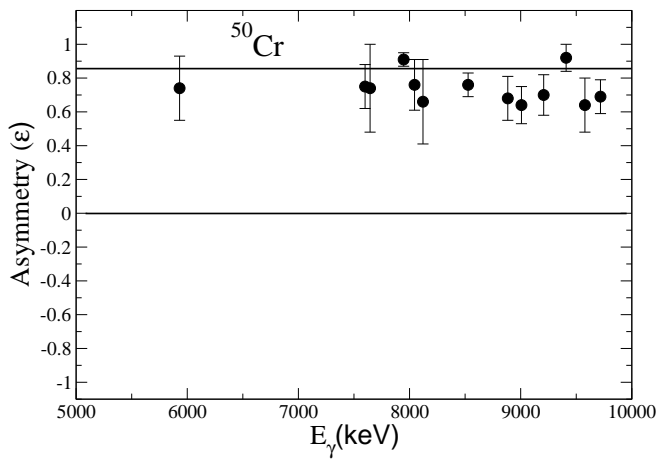}
    \caption{Experimental asymmetry for the $M1$ transitions observed
      at HI$\gamma$S. \label{as}}
  \end{center}
\end{figure}

Parity quantum numbers of spin-1 states were determined at the High
Intensity $\gamma$-Ray Source~\cite{higs1} operated by the
Triangle Universities Nuclear Laboratory and the Duke Free Electron
Laser Laboratory (DFELL) at Duke University in Durham, NC, USA.
This facility provides a quasi-monoenergetic beam of nearly completely
linearly polarized photons from Laser Compton Backscattering (LCB) in the entrance channel.
For details of the analysis procedure, we refer the reader to Refs.~\cite{nph,npm}.
The measurement was performed with 2.0 g of an isotopically enriched
$^{50}$Cr target (96.416\% enriched). The target was centered
between four HPGe detectors which were placed in a cross-like geometry
at polar angles of $90^{\circ}$
with respect to the incident beam, two of them have 100\% relative
efficiency and the other two have 60\% relative efficiency, respectively.
In this setup, two detectors were placed in the (horizontal) polarization plane of the incident $\gamma$-ray beam at azimuthal
angles $\varphi = 0^{\circ}$ and $\varphi = 180^{\circ}$ and the other two were placed perpendicular to it
at $\varphi = 90^{\circ}$ and $\varphi = 270^{\circ}$, respectively.

The experimental asymmetries $\varepsilon$ for the $M1$ transitions observed in the measurements at HI$\gamma$S are shown
in Fig.~\ref{as}. The asymmetry $\varepsilon$ is defined by the ratio of the measured and efficiency-corrected peak
intensities $A_{\parallel}$ and $A_{\perp}$, within and perpendicular to the polarization plane, respectively:
\begin{equation}
  \varepsilon = \frac{A_{\parallel}-A_{\perp}}{A_{\parallel}+A_{\perp}} = q \cdot \Sigma.
  \label{eq_asymmetry}
\end{equation}
Here, $\Sigma$ is the analyzing power and $q$ is the experimental sensitivity which accounts for the finite opening angles of the detectors
and the finite size of the target. The analyzing power $\Sigma$ of NRF amounts to $+1$ for a $J^{\pi}=1^+$ state and
$-1$ for $J^{\pi}=1^{-}$ state, respectively.
For this setup, the experimental sensitivity $q$ amounts to 0.86(1). In total 13 $M1$ excitations of $^{50}$Cr have been observed in this experiment.

\begin{table*}[htb]
\caption{\label{tab:Table1} Transitions observed in the
  $^{50}$Cr($\gamma$,$\gamma^\prime$) reaction with an endpoint
  energy of 9.7 MeV. Experimental uncertainties of the excitation
  energies are less then 0.5 keV.   \label{tab_1}}
\begin{ruledtabular}
  \begin{tabular}{ccccccccccc}
%\begin{longtable}{ccccccccccc}
$E_{x}$$^{\footnotemark[1]}$~~ &~~$E_{\gamma}$$^{\footnotemark[1]}$~~&~~ $\frac{W(90^\circ)}{W(130^\circ)}$~~ &$J^{\pi}$ &~~ $I_{i,0}$
~~~ & $\Gamma_{0}$~&$\frac{\Gamma_{i}}{\Gamma_{0}}$$^{\footnotemark[2]}$~&~ $B(M1)\uparrow$~&~$B(M1)\uparrow$~ &$ {\Gamma_0}$$^{red}$ \\
(keV) & (keV)&&& (eVb) & (eV) && ($\mu^{2}_{N}$) &($\mu^{2}_{N}$) &(meV/$\rm MeV^{3}$) \\
\hline
 3628.2 &3628.0 &0.74(2) & $1^{+}$${\footnotemark[3]}$&120(5)&0.205(9)&&1.113(49)&0.99(12)$^{\footnotemark[4]}$&4.29(19) \\

    &2845.0&&&&&0.49(1)&&&\\

4997.0 &4996.7 &0.78(6) & $1^{(+)}$${\footnotemark[5]}$&16.2(12)&0.070(7)&&0.145(15)&-&0.56(6) \\

    &4213.8&&&&&1.0(1)&&&\\

% 5362.0 &5361.7 &1.19(23) & $1^{\pm}$&4.4(4)&0.011(1)&-&0.018(2)&-&0.205(19)&0.07(1)\\

% 5528.3 &5528.0 &0.93(10) & $1^{\pm}$&31.5(12)&0.084(3)&-&0.129(5)&-&1.425(51)&0.50(2)\\

 %5783.7 &5783.4 &0.93(15) & $1^{\pm}$&13.2(8)&0.038(2)&-&0.051(3)&-&0.563(30)&0.20(1)\\

 5931.2 &5930.8 &0.92(9) & $1^{+}$&23.9(20)&0.073(6)&-&0.091(7)&-&0.35(3)\\

% 6817.5 &6817.0 &0.75(18) & $1^{\pm}$&14.8(15)&0.060(6)&-&0.049(5)&-&0.543(54)&0.19(2)\\

 %6312.6 &6312.2 &0.87(14) & $1^{-}$&11.4(9)&0.040(3)&-&-&-&0.456(34)&0.16(1)\\

 %6504.4 &6503.9 &0.60(8) & $1^{-}$&82.7(41)&0.305(15)&-&-&-&3.18(16)&1.11(5)\\

 %6897.8 &6897.3 &0.89(15) & $1^{-}$&41.5(20)&0.172(8)&-&-&-&1.503(70)&0.52(2)\\

 %6932.4 &6931.9 &0.76(4) & $1^{-}$&130.8(32)&0.547(13)&-&-&-&4.71(11)&1.64(4)\\

 %7104.0 &7103.4 &0.71(4) & $1^{-}$&103.2(27)&0.453(12)&-&-&-&3.622(96)&1.26(3)\\

 7600.8 &7600.2 &0.80(10) & $1^{+}$&66.4(73)&0.334(37)&-&0.197(22)&-&0.76(8)\\

 7645.7 &7645.1 &0.69(17) & $1^{+}$&23.2(28)&0.118(14)&-&0.068(8)&-&0.26(3)\\

 %7840.3 &7839.6 &0.67(16) & $1^{-}$&18.5(21)&0.099(11)&-&-&-&0.589(65)&0.21(2)\\

 %7880.9 &7880.2 &0.65(5) & $1^{-}$&68.6(27)&0.371(15)&-&-&-&2.17(9)&0.76(3)\\

 7948.1 &7947.4 &0.78(3) & $1^{+}$&197.7(96)&1.382(79)&&0.714(41)&-&2.75(16)\\

         &7164.5&&&&&0.27(2)&&&\\

 8045.8 &8045.1 &0.77(10) & $1^{+}$&42.2(47)&0.238(26)&-&0.118(13)&-&0.46(5)\\

 %8111.2 &8110.5&0.67(4) & $1^{-}$&213.5(65)&1.552(64)&0.27(2)&-&-& 8.34(34)&2.91(12)\\

    %    &7326.6&&&&&&&&\\

 %8153.7 &8153.0&0.70(13)& $1^{\pm}$&30.9(25)&0.179(14)&-&0.086(7)&-&0.947(74)&0.33(3)\\

 8121.5 &8120.8 &0.76(20) & $1^{+}$&16.4(20)&0.094(11)&-&0.045(5)&-&0.18(2)\\

 %8359.4 &8358.6&0.65(10)&$1^{-}$&41.9(26)&0.255(16)&-&-&-&1.252(79)&0.44(3)\\

 %8439.6 &8438.8&0.85(8)&$1^{-}$&81.1(34)&0.502(21)&-&-&-&2.39(10)&0.84(3)\\

 8528.1 &8527.4&0.70(8)&$1^{+}$&96(11)&0.85(11)&&0.353(48)&-&1.36(18)\\

        &7743.1&&&&&0.39(6)&&&\\

 %8651.3 &8650.5&0.86(7)&$1^{-}$&122.2(41)&0.796(27)&-&-&-&3.52(12)&1.23(4)\\

 %8714.7 &8713.9&0.89(8)&$1^{-}$&99.6(45)&0.981(60)&0.49(4)&-&-&4.25(26)&1.48(9)\\

  %      &7930.9&&&&&&&&\\

 %8770.3 &8769.4&0.95(14)&$1^{-}$&37.8(28)&0.433(45)&0.71(9)&-&-&1.84(19)&0.64(7)\\

   %    &7985.7&&&&&&&&\\

 8885.6 &8884.8&0.82(8)&$1^{+}$&77.7(68)&0.534(47)&-&0.197(17)&0.277(74)${\footnotemark[5]}$&0.76(7)\\

 %8998.3 &8997.4&0.79(13)&$1^{-}$&40.0(31)&0.282(22)&-&-&-&1.110(87)&0.39(3)\\

 9007.9 &9007.0&0.88(13)&$1^{+}$&40.5(48)&0.286(34)&-&0.101(12)&-&0.39(5)\\

 %9087.6 &9086.7&0.85(15)&$1^{-}$&102.8(52)&0.739(37)&-&-&-&2.82(14)&0.99(5)\\

 %9174.0 &9173.1&0.61(14)&$1^{-}$&48.2(44)&0.353(32)&-&-&-&1.31(11)&0.46(4)\\

 9208.3 &9207.4&0.77(20)&$1^{+}$&50(12)&0.369(89)&-&0.123(30)&-&0.47(11)\\

 9409.5 &9408.5&0.78(14)&$1^{+}$&105(17)&0.81(13)&-&0.252(41)&-&0.97(16)\\

 %9552.6 &9551.7&0.79(11)&$1^{-}$&108.1(71)&0.858(56)&-&-&-&2.82(18)& 0.98(6)\\

 9579.1 &9578.1&0.97(25)&$1^{+}$&37.7(78)&0.301(62)&-&0.089(18)&-&0.34(7)\\

 9719.1 &9718.1&0.85(11)&$1^{+}$&173(21)&1.42(17)&-&0.402(49)&-&1.55(19)\\
  \end{tabular}
%\end{longtable}
\footnotetext[1]{The difference between the the excitation energy ($E_{x}$) and transition energy (gamma-ray energy ($E_{\gamma}$)) is due to the nuclear recoil.}
\footnotetext[2]{Branching ratio.}
\footnotetext[3]{From Refs.~\cite{ppr,hs,ar}.}  
\footnotetext[4]{From Ref.~\cite{en}.}
\footnotetext[5]{From Ref.~\cite{nds}.}
\end{ruledtabular}
\end{table*}
%%%%%%%%%%%%%%%%%%%%%%
% Figure 3
%%%%%%%%%%%%%%%%%%%%%
\begin{figure}[htb]
  \begin{center}
    \includegraphics*[width=\linewidth]{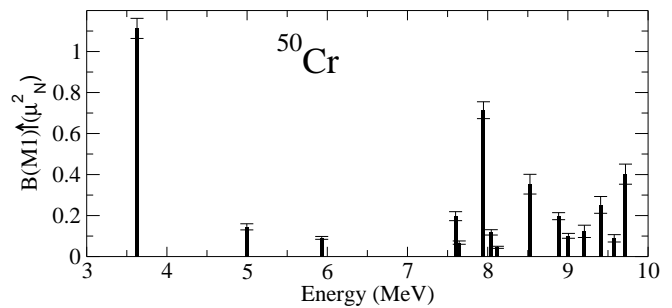}
    \caption{The experimental $B(M1)$ values for the
      ground state transitions.\label{bm1}}
  \end{center}
\end{figure}

\section{Results and discussion}
\subsection{Experimental results}

The experimental results are summarized in Table~\ref{tab_1} and Figure~\ref{bm1}.
In the present work, we have deduced $M1$ strengths in $^{50}$Cr
for the first time through the ($\gamma$,$\gamma^\prime$) reaction
in addition to the 3628.0-keV transition, whose $M1$ strength has previously been measured
via ($\gamma$,$\gamma^\prime$)~\cite{en} and ($e$,$e^\prime$)~\cite{ppr,hs,ar}
experiments albeit with larger uncertainties.
In the present experiment, the uncertainty of the cross
section has been reduced for the 3628.0-keV transition.
%to about 2\%.
Nine of the observed excited states of $^{50}$Cr may be
already known from a ($p$,$p^\prime$) experiment~\cite{ppr} and
three states (3628.2, 7645.7 and 8885.6 keV) were known
from ($\gamma$,$\gamma^\prime$)~\cite{en,rm} experiments. For twelve
$1^{+}$ states of $^{50}$Cr, our data provide first information on
their $\gamma$-decay transitions.
%%%%%%%%%%%%%%%%%%%%%%
% Figure 4
%%%%%%%%%%%%%%%%%%%%%
\begin{figure}[htb]
  \begin{center}
    % \vspace{0.3cm}
    \includegraphics*[width=6.5cm]{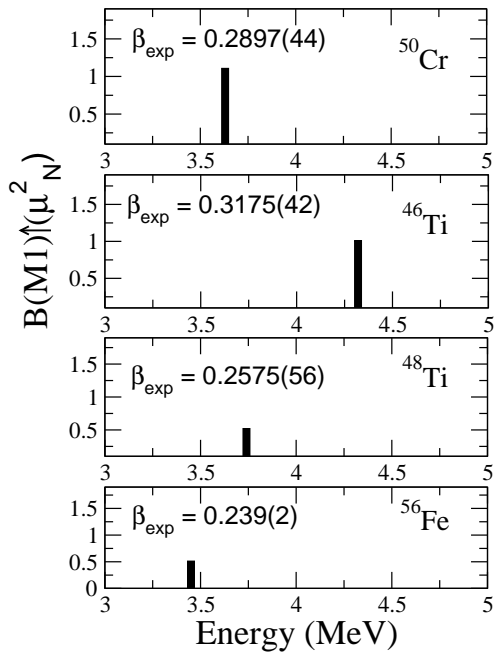}
    \caption{B(M1)$\uparrow$ values for $1^{+}_{sc}$ states of
      $^{46,48}$Ti~\cite{ar} and $^{56}$Fe~\cite{fe} along with the 3.628
      MeV state of $^{50}$Cr. The experimental deformation parameters
      $\beta_{\rm{exp}}$ \protect\cite{bnl} are given for each
      nucleus. \label{ss}}
  \end{center}
\end{figure}

Using linearly polarized quasi-monoenergetic photons at HI$\gamma$S, positive
parity quantum numbers were assigned to 13 states apart from the
3628.2-keV state. The definite positive parity quantum number assignment for the 3628.2-keV state has
been adopted from earlier ($p$,$p^\prime$ ) and
($e$,$e^\prime$)~\cite{ppr,hs,ar} experiments. Three new $M1$ transitions
(5930.8, 8045.1 and 8120.8 keV) have been identified in
the present work and for thirteen $1^{+}$ states $M1$ excitation strengths have been
obtained for the first time.
%%%%%%%%%%%%%%%%%%%%%%
% Figure 5
%%%%%%%%%%%%%%%%%%%%%
\begin{figure}[htb]
  \begin{center}
    \includegraphics[width=\linewidth,height=7.5cm]{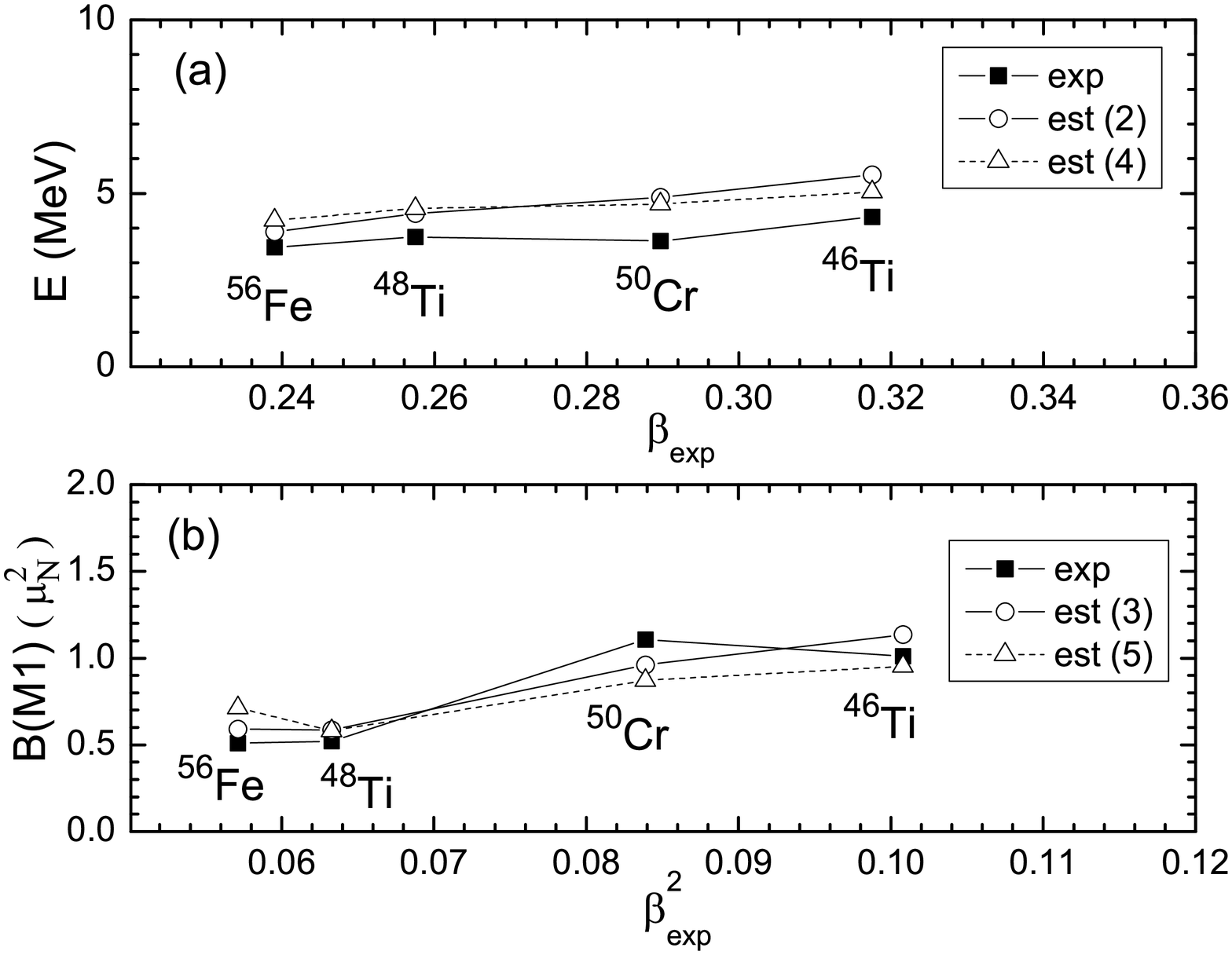}
    \caption{Experimental $1^{+}_{sc}$ energies (top) and B(M1)-values (bottom) as
      compared to  estimations
(\ref{1E})-(\ref{1BM1}) and (\ref{2E})-(\ref{2BM1})
%      \protect\ref{2} and \protect\ref{3}
in $^{46,48}$Ti~\cite{ar}, $^{50}$Cr (present work)  and
      $^{56}$Fe~\cite{fe}. The experimental deformation parameters
      $\beta_{\rm{exp}}$ are taken from \protect\cite{bnl}.
      \label{sys}}
  \end{center}
\end{figure}

The experimental $B(M1)$ values into the ground state are shown in
Figure~\ref{bm1}. The $M1$ strength distribution exhibits an
intriguing pattern: an isolated, rather strong $M1$ excitation at low energy
at 3.6 MeV is separated from an accumulation of fragmented $M1$
strength above 7 MeV. A strongly excited $J^{\pi}$ = $1^{+}$ state at an
excitation energy of 3628.2 keV with a strength of $B(M1)\uparrow$ =
1.113(49) $\mu^{2}_{N}$ is observed. The rest of the observed $M1$
excitations in $^{50}$Cr is expected~\cite{ppr} to be
generated by spin excitations, mainly from
$f_{7/2} \rightarrow f_{5/2}$ orbitals like in $^{52}$Cr~\cite{pai}.
We will first concentrate on the $1^{+}$ state at 3628.2 keV.
This state
could correspond to a moderately collective scissors-like mode.
This mode has been
predicted~\cite{Iudice,ic} and then discovered in ($e$,$e^\prime$) reactions by A. Richter
 {\it et al.}~\cite{db} in heavy deformed
nuclei. As mentioned above, the $1^{+}_{sc}$ state was also reported in the open $fp$ shell
nuclei $^{46}$Ti, $^{48}$Ti~\cite{ar} and $^{56}$Fe (Z=26) \cite{fe}.

The experimental value of the quadrupole deformation parameter $\beta_{\rm{exp}}$ for
$^{50}$Cr is 0.2897(44)~\cite{bnl}, which indicates that this nucleus is
well deformed and hence can support the formation of a $1^{+}_{sc}$ state.
The ratio of the $M1$ transition rates of the $1^{+}$ level at 3628.2 keV
to the ground state and the first-excited $2^{+}$ state (at 783.32 keV excitation
energy) is found to be
$\frac {B(M1;1^{+}_{1} \to 2^{+}_{1})} {B(M1;1^{+}_{1} \to 0^{+}_{1})}$ =
$\frac {\Gamma_{1}}{\Gamma_{0}}\cdot$$(\frac {E_{\gamma0}}{E_{\gamma1}})^3$
= 1.02(2) with $E_{\gamma0,(1)}$ being the $\gamma$-ray transition energy to the
ground (first-excited) state.
This value is larger by a factor of 2 compared to the value
of 0.5 obtained by employing the Alaga rule~\cite{al}.
This discrepancy hints at some degree of K mixing, perhaps $^{50}$Cr
exhibits some degree of gamma-softness \cite{sh1}.

Since in $^{50}$Cr neutron and proton Fermi levels are in the
same open subshell, spin magnetism is also expected in the lowest
$1^{+}$ state. Therefore, the state at 3628.2 keV should be of a mixed
orbital/spin nature. It is also excited in ($p$,$p^\prime$) reactions~\cite{ppr},
which points to a sufficient contribution of spin-$M1$ strength.

The available experimental results for the $1^{+}_{sc}$ in $fp$-shell nuclei,
including the present data, are exhibited in Fig.~\ref{ss}. It is
instructive to see how much these data match early
\cite{Har01,Hei10,NLI97,Iud_Rich}
\begin{eqnarray}
  E&=&66 \delta A^{-1/3} \rm{MeV} ,
  \label{1E}
  \\
  B(M1)&=&0.0042 E A^{5/3} \delta^2 g_r^2 \; \mu^2_N ,
  \label{1BM1}
\end{eqnarray}
and modified \cite{Pie98}
\begin{eqnarray} \label{2E}
  E &=&13.4 \sqrt{1+(3\delta)^2} A^{-1/3} \rm{MeV} ,
%\end{equation}
% \cite{Hei10,NLI97,Iud_Rich},
%\begin{equation}
\\
\label{2BM1}
  B(M1)&=&0.66 \frac{\delta^3}{1+ (3\delta)^2} A^{4/3} g_{\rm{eff}}^2 \; \mu^2_N ,
\end{eqnarray}
empirical estimations for
$1^{+}_{sc}$ energies and $B(M1)$ values.
Here, $\delta = 0.946 \beta$ is the deformation parameter; $g_r= 2Z/A$ is the
%normalized isovector
orbital factor, $g_{\rm{eff}}=c_g g_r$ with $c_g$=0.8 \cite{Pie98}.
The deformation parameter
$\beta$ in Figs. 4-7 is related to the value $B(E2)\uparrow =B(E2,0^{+}_{1} \to 2^{+}_{1})$
%$Q_{2} = \int d{\mathbf{r}} \rho_p({\mathbf{r}}) r^2 Y_{20}$
as \cite{Raman}
$\beta_{\rm{exp}}  = \frac{4\pi}{3ZR^2_0} \sqrt{B(E2)\uparrow/e^2}$
with $R_0 = 1.2\; A^{1/3}\rm{fm}$.
As seen below from Fig.~\ref{sys}, estimations (\ref{1E})-(\ref{1BM1}) and (\ref{2E})-(\ref{2BM1})
give for $fp$-shell nuclei rather close results.

The data are compared to the estimates according to Eqs.
(\ref{1E})-(\ref{1BM1}) and (\ref{2E})-(\ref{2BM1}) in Fig.~\ref{sys}.
We see in panel (a) that the experimental values lie somewhat below the
estimation but closely
follow the linear dependence on the deformation parameter, which is
a strong fingerprint of a scissors-like excitation.
Our data on $^{50}$Cr well fit this trend. Following the panel (b), the correspondence of
the experimental and estimated $B(M1)$-values is also quite satisfactory.
%Such a deviation may not be considered a severe discrepancy from the
Some deviations from the phenomenological estimates can be explained by
that in $fp$-nuclei the states are determined by a few two quasiparticle (2qp)
components and the orbital motion is supplemented by a significant spin admixture (see discussion below). Altogether one may state that our results for the
state at 3.628 MeV rather well fit the $1^{+}_{sc}$ systematics, which justifies the assignment of predominantly scissors-like nature to this state. In the
next section, our theoretical analysis presents  additional arguments in favor of this conclusion.

\subsection{Theoretical analysis}

The $M1$ strength in $^{50}$Cr was analyzed within the
%the Quasiparticle-Phonon Model (QPM~\cite{qpm1,qpm2,qpm3}),
Skyrme  Quasiparticle Random-Phase-Approximation (QRPA)
\cite{Ring,Repko} and a Large-Scale Shell Model (LSSM) \cite{lssm_Gaurier}.

\subsubsection{Skyrme QRPA}

\begin{figure}[htb]
  \begin{center}
    \includegraphics*[width=\linewidth,height=6.5cm]{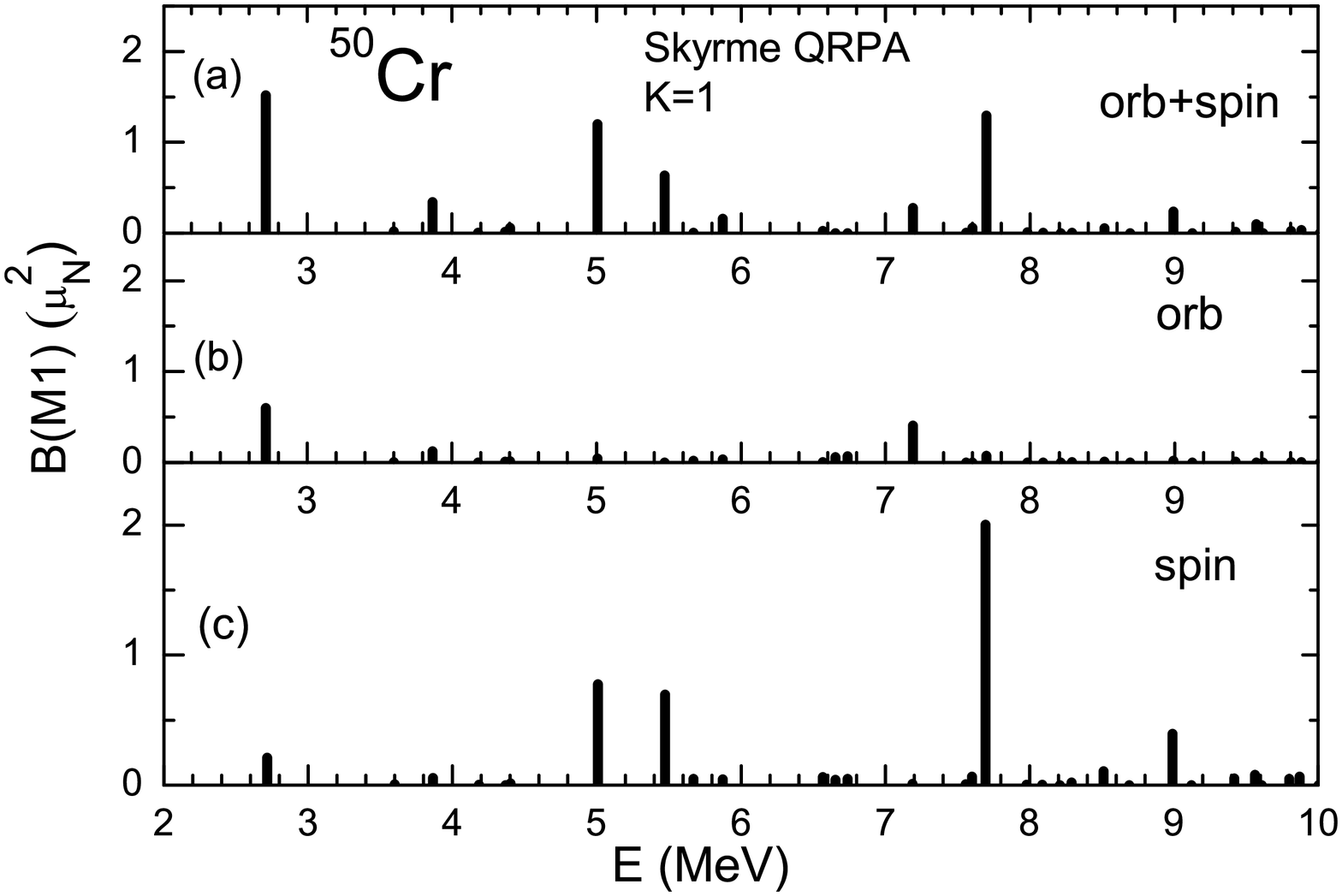}
    \caption{Total (a), orbital (b), and spin (c)
      M1 strengths K=1 calculated within Skyrme QRPA with the force
      SGII. \label{rpa1}}
  \end{center}
\end{figure}

\begin{figure}[htb]
  \begin{center}
    \includegraphics*[width=\linewidth,height=6.3cm]{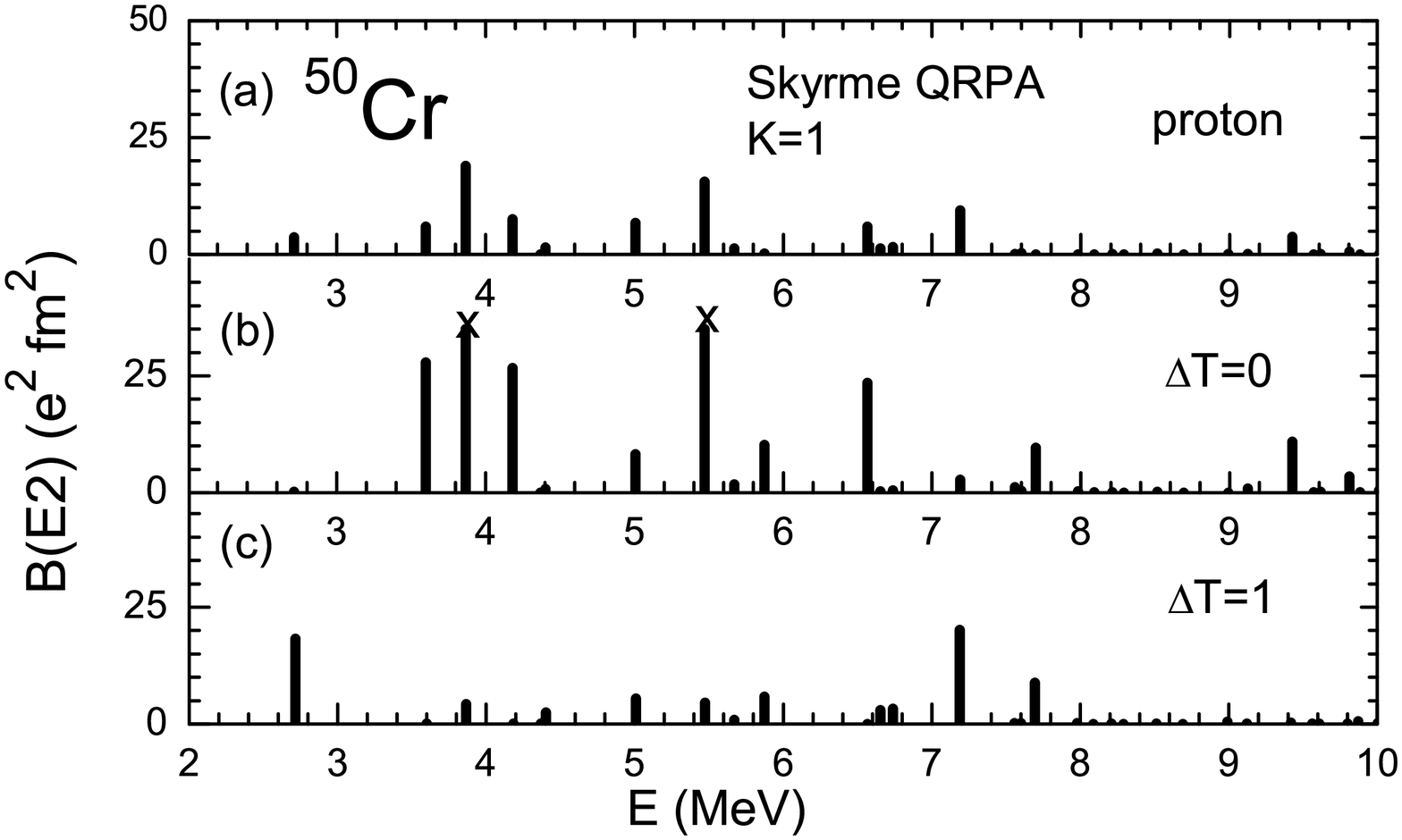}
    \caption{Proton (a), isoscalar (b), and isovector (c) E2
      strengths K=1 calculated within the Skyrme QRPA with the force SGII. In
      panel (b), high peaks at 3.9 and 5.5 MeV (with B(E2)=116 and
      101 $\rm{e}^2 \rm{fm}^4$, respectively) are marked by
      crosses. \label{rpa2}}
  \end{center}
\end{figure}

\begin{figure}[htb]
  \begin{center}
    \includegraphics*[width=\linewidth]{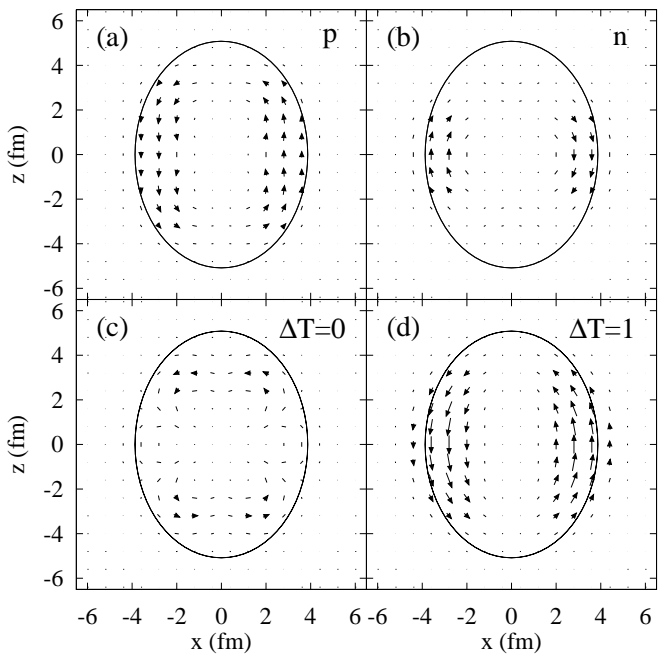}
    \caption{Proton (a), neutron (b), isoscalar (c), and isovector (d)
      orbital currents in x-z plane for the 2.71 MeV-state calculated
      in the Skyrme QRPA with the force SGII. In all the panels, the currents are
      equally scaled to demonstrate the most strong flows. The magnitude of the
      current is determined by the arrow length.
      \label{rpa3}}
  \end{center}
\end{figure}

\begin{figure}[htb]
  \begin{center}
    \includegraphics*[width=\linewidth,height=6.5cm]{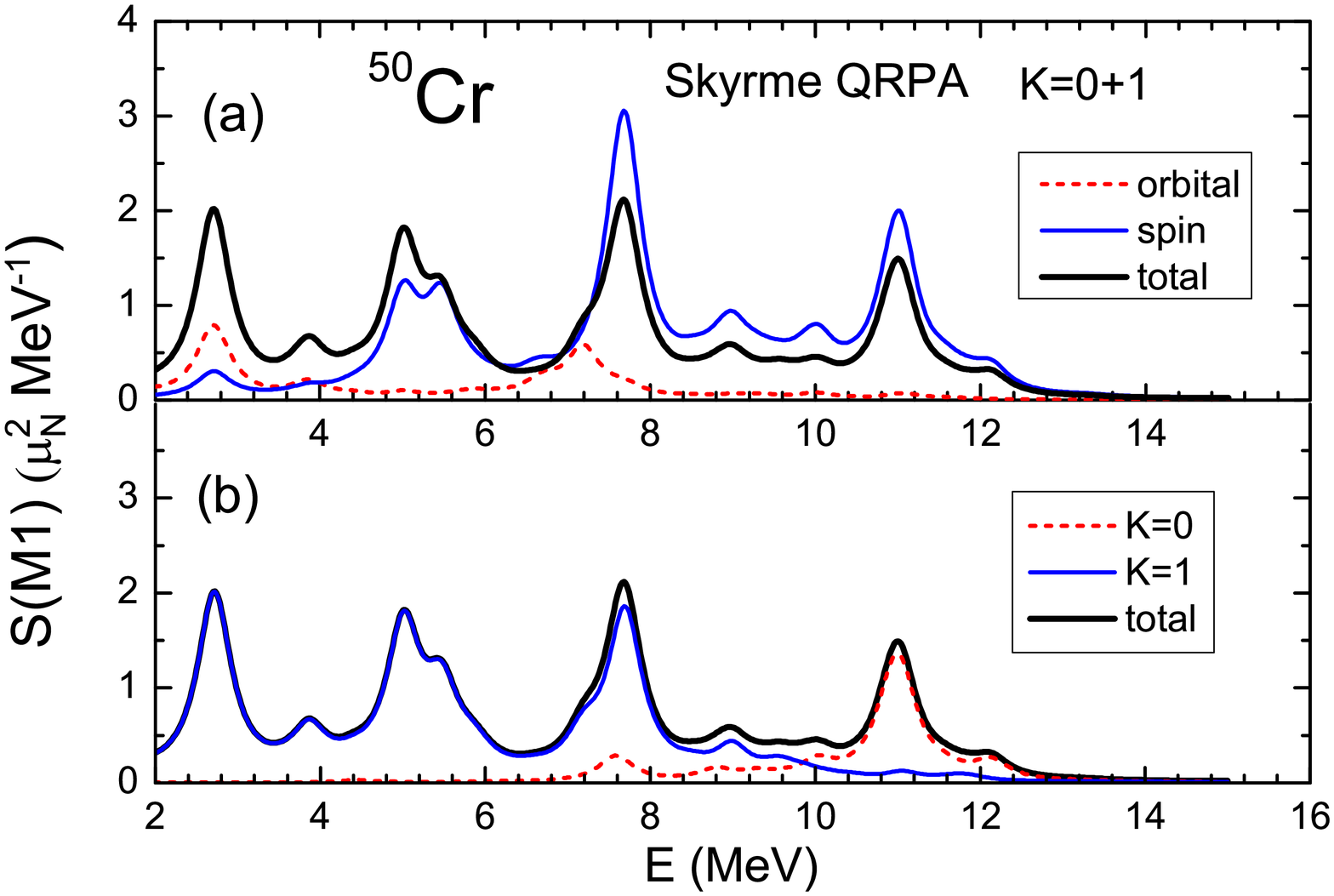}
    \caption{(Color online) Strength functions in $^{50}$Cr smoothed by the
    Lorentz weight with $\Delta$=0.5 MeV.
    Panel (a): orbital (red dashed curve), spin (blue solid curve)
    and total orbital+spin  (black bold curve)
    strengths. Contributions from all the projections (K=0 and 1) are included.
    Panel (b): K=0 (red dashed curve), K=1  (blue solid curve) and
    total (black bold curve) strengths.}
    \label{sf}
  \end{center}
\end{figure}

Results of the self-consistent Skyrme-QRPA calculations are presented in
Figs.~\ref{rpa1}-\ref{sf} and Tables II and III. The calculations are performed
with a QRPA restricted to axial symmetry \cite{Repko} for axially deformed
nuclei. The Skyrme QRPA method is fully self-consistent
since i) both the mean field and the residual interaction are derived from the
same Skyrme functional and ii) the residual
interaction includes all the functional contributions as well as the Coulomb
direct and exchange terms. The Skyrme
parameterization SGII \cite{SGII} is used. The volume
$\delta$-force pairing is treated at the BCS level \cite{Ben00}. The parameter of
the equilibrium axial quadrupole deformation $\beta$=0.30 is determined by minimization
of the total energy of the system. The calculated $\beta$ is in
a satisfactory agreement with the experimental value $\beta_{\rm{exp}}$=0.2897(44) \cite{bnl}.
The spurious strength is concentrated below 2 MeV (not shown in Figs.~\ref{rpa1},~\ref{rpa2} and \ref{sf})
and does not noticeably affect the results. Note that applications of the self-consistent
Skyrme QRPA to magnetic  orbital and spin excitations in deformed nuclei are still
very limited and mainly devoted to heavy deformed nuclei, see e.g.
\cite{Nest_M1,Nest_M1_Nd}. To our knowledge, the present Skyrme QRPA exploration is the first one
for medium deformed nuclei.
%%%%%%%%%%%%%%%%%%%%%%
% Figure 6
%%%%%%%%%%%%%%%%%%%%%

Figure~\ref{rpa1} presents calculated $B(M1)$ values
obtained for the total (orbital + spin), orbital, and spin
$M1$ transitions $I^{\pi}_{K}=0^+_{0,{gs}} \to 1^+_{1,\nu}$ from the ground state
to the $\nu$-th QRPA state with $K^{\pi}=1^+$.  In the spin part,
the familiar quenching factor of 0.7 is used. The calculations result in a $1^+$ state at
2.71 MeV with a large $M1$ excitation strength of $B(M1)$$\uparrow$=1.52 $\mu_N^2$.
By comparing $B(M1)$
values in the plots a), b) and c), it is easy to see that the magnetic
strength of this state is produced by a constructive interference of
the orbital and spin parts. The orbital strength dominates which
indicates a predominantly scissors-like character of the state.
The mainly scissors-like nature of this state is also
supported by a satisfactory agreement of its excitation energy and
$B(M1)$ strength with the systematics for the $1^{+}_{sc}$ as a function of
nuclear deformation and mass number. Note that the calculated 2.71-MeV
state is not purely orbital but has some spin admixture, which is
typical~\cite{li} for light and medium deformed nuclei. Despite a noticeable energy
difference, the 2.71-MeV state corresponds to the experimentally observed
state at 3.628 MeV because both of them deliver  experimentally
and theoretically a maximal $B(M1)$ strength
in the energy range 2-4 MeV.
%%%%%%%%%%%%%%%%%%%%%%
% Figure 7
%%%%%%%%%%%%%%%%%%%%%

The scissors-like character of the calculated 2.71-MeV, $1^{+}$ state can be additionally
confirmed by comparison of the proton, isoscalar ($\Delta$T=0) and isovector
($\Delta$T=1), $\Delta$K=1 B(E2;$I^{\pi}_{K}=0^{+}_{0,gs}\to 2^+_{1,\nu}$)
%_{K}=0^{+}_{0,gs}) %\to 2^+_{1,$\nu$}) 
values of the associated 2$^+_{1,\nu}$ state belonging to the same band. These values were obtained with the effective charges
$e_{\rm{eff}}^p=1, e_{\rm{eff}}^n=0$;
$e_{\rm{eff}}^p=e_{\rm{eff}}^n=1$; and
$e_{\rm{eff}}^p=-e_{\rm{eff}}^n=1$, respectively. 
The B(E2) strengths are exhibited in  Fig.~\ref{rpa2}. For the convenience of
comparison with Fig.~\ref{rpa1}, the strengths are plotted as a function
of the excitation energies $E_{1^{+}_{1,\nu}}$
for the QRPA states with K=1 (The energies of ($0^{+}_{0,gs}\to 2^+_{1,\nu}$)
transitions can be roughly estimated in the approximation that the
moment of inertia of K=1 bands is the same as in the ground-state band.
Then the rotational  increment ($E_{2^{+}_{1,\nu}}$-$E_{1^{+}_{1,\nu}}$) to $E_{1^{+}_{1,\nu}}$ is  $\sim$522 keV.).
Fig. ~\ref{rpa2}  shows that, for the band based on
the calculated 2.71-MeV state, we get a negligible B(E2, $\Delta$T=0) but
sizeable B(E2, $\Delta$T=1) value. This clearly emphasizes the
isovector excitation character of the state.
%Following
%Fig.~\ref{rpa2}, the calculated 2.71-MeV state has a negligible B(E2, $\Delta$T=0) but a
%sizeable B(E2, $\Delta$T=1) value. This fact clearly emphasizes its isovector excitation character.
Moreover, the B(E2, $\Delta$T=1) value calculated for this state is the largest for all states below 6 MeV.
This observation points to the close relation of this state to the underlying quadrupole deformation.
Both aspects match the isovector nature of the scissors-like mode which is usually
viewed as rotation-like out-of-phase oscillations of the deformed proton and
neutron subsystems.

The nature of the 2.71-MeV state is further
inspected in Fig.~\ref{rpa3}, where distributions (proton, neutron,
isoscalar and isovector)  of the orbital transition current of this state are exhibited.
The distributions represent the proton (p) and  neutron (n) transition densities of the
 convection current, as well as their sums ($\Delta$T=0) and differences ($\Delta$T=1). For the
 convenience of the view, the currents in all the panels are equally scaled so as
 to exhibit mainly the strong flows.
The figure shows that the orbital motion is basically represented by rotation-like
out-of-phase oscillations of protons and neutrons. Indeed in the regions of most intense
flow (at the left and right nuclear surface), the protons and neutrons move in
the opposite directions (compare plots a) and b)).
In these regions, the $\Delta$T=1 motion dominates over the $\Delta$T=0 one. It is remarkable that
the $\Delta$T=1 motion is most pronounced in the region of the nuclear equator and weak at
the nuclear poles. This is surprising since the scissors-like picture suggests the
opposite. One may suggest that the orbital motion in 2.71-MeV state actually
corresponds not to the scissors-like flow proposed by N. Lo Iudice and F. Palumbo
\cite{Iudice} but to an alternative picture from E. Lipparini and S. Stringari
\cite{lip_str_83}, where the isovector motion takes place inside the rigid surface
and is maximal in the equator region. These two pictures are analogous to the
Goldhaber-Teller \cite{GT} and Steinwedel-Jensen \cite{SJ} treatments of the
 electric isovector giant dipole resonance, respectively. Our self-consistent
Skyrme-QRPA results more closely correspond to the Lipparini-Stringari picture for the
low-energy orbital $M1$ mode. 
%Note that as compared to \cite{Iudice}, the
%Lipparini-Stringari estimations takes into account the coupling of the scissors mode
%and K=1 part of the isovector giant quadrupole resonance.

Now let's briefly discuss the states at $E >$ 4 MeV, i.e. above the scissors-like mode.
As seen from Fig.~\ref{rpa1},  these states are
mainly of a spin nature with only a few exceptions at $\sim$4  and 6.5-7.5
MeV. The spin states form two groups at $\sim$5 and $\sim$8 MeV, that
constitute altogether the K=1 branch (with both K=$\pm$1 contributions)
of the $M1$ spin giant resonance.
%%%%%%%%%%%%%%%%%%%%%%
% Figure 8
%%%%%%%%%%%%%%%%%%%%%

A more general view of this resonance is given
in Fig.~\ref{sf} where the M1 strength function
%\begin{equation}\label{M1_sf}
%S(M1; E) = \sum_{I^{\pi}_{k}} |\:\langle I^{\pi}_{k} |
%\hat{M}(M1) | 0 \:\rangle |^2 \: \xi_{\Delta}(E-E_{I^{\pi}_{k}} )
%\end{equation}
%is presented. Here $|0\rangle$ is the ground state wave function, $|I^{\pi}_{k}\rangle$ and
%$E_{I^{\pi}_{k}}$ are QRPA states and energies,
%$\hat{M} (M1)$ is the M1
%transition operator, $\xi_{\Delta}(E-E_{I^{\pi}_{k}}) = \Delta /(2\pi
%[(E-E_{I^{\pi}_{k}})^2 - \Delta^2/4]$ is the Lorentz smoothing with the averaging
%parameter $\Delta$=0.5 MeV.
\begin{equation}\label{M1_sf}
S(M1; E) = \sum_{\nu} |\:\langle 1^{+}_{K,\nu} |
\hat{M}(M1,K) | 0 \:\rangle |^2 \: \xi_{\Delta}(E-E_{1^{+}_{K,\nu}} )
\end{equation}
is presented. Here $|0\rangle$ is the ground state
wave function, $|1^{+}_{K,\nu}\rangle$ and
$E_{1^{+}_{K,\nu}}$ are QRPA states and energies,
$\hat{M} (M1,K)$ is the M1 transition operator with
the projection K, $\xi_{\Delta}(E-E_{1^{+}_{K,\nu}}) = \Delta /(2\pi
[(E-E_{1^{+}_{K, \nu}})^2 - \Delta^2/4]$ is the Lorentz smoothing with the averaging
parameter $\Delta$=0.5 MeV.

Figure~\ref{sf} shows that the spin resonance basically consists
of two bumps at $\sim$5 and $\sim$8 MeV
from the K=1 branch and the bump at $\sim$11 MeV from the branch K=0. Altogether spin excitations
fill a broad region from 5-13 MeV, which well matches a resonance region in LSSM calculations,
see Fig.~\ref{shell} below. In our QRPA calculations, the spin resonance is not concentrated in one broad bump
but looks like a sequence of few well separated structures. This may be expected from the fact that QRPA omits
parts of the spin correlations. As seen from Fig.~\ref{shell}, the inclusion of more correlations in the LSSM allows
to mix the separate structures and to form a broad regular resonance.

Figure~\ref{sf}a) shows that the interference of the
orbital and spin excitations is constructive at E$<$7 MeV and destructive at higher energies. As
seen from Fig.~\ref{sf}b), the excitations in the energy interval of our interest, E$<$10 MeV,
almost completely belong to the K=1 branch. In the energy interval 4.5-14 MeV where the spin M1 resonance is localized,
the calculated total M1 strength is $\sim$5.9 $\mu_N^2$. This is in accordance with the experimental strengths
of 5-7  $\mu_N^2$ found in $^{24}$Mg and $^{28}$Si \cite{Har01}. For the interval 2-9.7 MeV covered by the present
experiment, the calculations give a total strength of $\sim$6.2 $\mu_N^2$
which somewhat overestimates the total observed $M1$ strength $\sim$ 4.01 $\mu_N^2$.
%to the experimental total strength $\sim$3.86 $\mu_N^2$.
%%%%%%%%%%%%%%%%%%%%%%
% Figure 9
%%%%%%%%%%%%%%%%%%%%%
\begin{table}
\caption{\label{tab:Table2} The energies $E$, $B(M1)$ values and two maximal
2qp components of particular $I^{\pi}_{K,\nu}$ QRPA states
with K=1 and 0.
The components are given in Nilsson quantum numbers $[Nn_z\Lambda]$.
$\mathfrak{N}$ denotes the contribution of the component to the state norm.}
\begin{ruledtabular}
\begin{tabular}{|c|c|c|c|c|}
 E  & B(M1)       & p/n & 2qp  & $\mathfrak{N}$ \\
 MeV & $\mu_N^2$  &     &            &      \\
\hline
 \multicolumn{5}{|c|}{K=1} \\
\hline
2.71 & 1.52 & n & [303]7/2 \; [312]5/2 & 69$\%$ \\
     &      & p & [312]5/2 \; [321]3/2 & 27$\%$ \\
5.01 & 1.20 & n & [321]3/2 \;[321]1/2 & 54$\%$ \\
     &      & p & [321]3/2 \; [321]1/2 & 33$\%$ \\
7.70 & 1.30 & n & [312]5/2 \;[301]3/2 & 40$\%$ \\
     &      & n & [321]3/2 \; [310]1/2 & 29$\%$ \\
\hline
 \multicolumn{5}{|c|}{K=0} \\
\hline
11.0 & 1.56 & p &  [312]5/2 \; [303]5/2 & 32$\%$\\
     &      & n & [321]3/2 \; [301]3/2 & 25$\%$ \\
\end{tabular}
\end{ruledtabular}
\end{table}

It is instructive to see the structure of the QRPA states providing
the dominant contribution to the orbital and spin bumps in Fig.~\ref{sf}.
This information is given in Table~\ref{tab:Table2}.
The table shows that the calculated $1^{+}_{sc}$ at 2.71 MeV
is formed by $\Delta$K=1 transitions between the
levels [303]7/2, [312]5/2 and [321]3/2. They are neighboring
levels arising due to the deformation splitting of the $1f_{7/2}$ subshell. Thus one
encounters here a typical scheme for orbital excitations. Note that the collectivity
of the 2.71-MeV state is low. Its two largest components exhaust 96$\%$ of the norm.
%At the same time, the energy of this state is upshifted by 0.38 MeV as compared
%to the energy of lowest 2qp state nn[303]7/2 [312]5/2. So the interaction effect is
%rather strong.

Table~\ref{tab:Table2} also shows the structure of the calculated states at 5.01,
7.70 and  11.0 MeV that strongly contribute to the main parts of the spin M1
resonance in Fig.~\ref{sf}. It is seen that these states are rather collective. Their
two largest components give altogether only 57-87$\%$ of the norm. Some of the components,
e.g.  [312]5/2-[301]3/2 in the 7.70 MeV-state and [321]3/2-[301]3/2,  [312]5/2-[303]5/2
in the 11.0 MeV-state
directly correspond to the $1f_{7/2}-1f_{5/2}$ spin-flip transition.
The strong deformation splitting spreads the strength of $1f_{7/2}-1f_{5/2}$ transitions
over different QRPA states. Since the QRPA description does not
include all correlations, the spin $M1$ resonance is yet represented by a small
number of states and looks like a sequence of a few separated structures.
The more comprehensive picture
of the spin $M1$ resonance is given below in terms of the shell model
which includes more correlations.
%%%%%%%%%%%%%%%%%%%%%%
% Figure 10
%%%%%%%%%%%%%%%%%%%%%
\begin{figure}[htb]
  \begin{center}
    \includegraphics*[width=\linewidth]{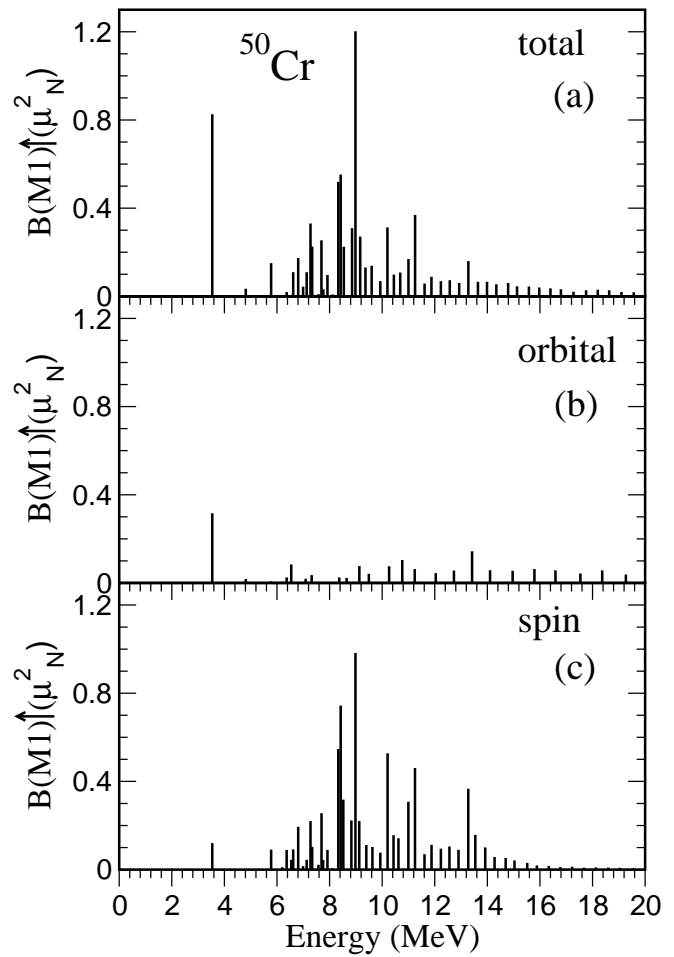}
%    \vspace{0.5cm}
    \caption{Total (a), orbital (b), and spin (c)
      M1 excitation strengths calculated within a large-scale shell-model with the
      KB3G interaction. \label{shell}}
  \end{center}
\end{figure}
\begin{table}
\caption{\label{tab:Table3}
Calculated $B(E2,0^{+}_{1} \to 2^{+}_{1})$$\uparrow$ value and characteristics of the lowest 1$^+_{1,\nu=1}$-state of $^{50}$Cr
(energy E, total excitation strength B(M1)$\uparrow$, orbital strength B(M1)$_{\rm{O}}$, spin strength B(M1)$_{\rm{S}}$, and ratio
R=B(M1)$_{\rm{O}}$/B(M1)$_{\rm{S}}$)
as compared to the present experimental data.}
%\set
\begin{ruledtabular}
\begin{tabular}{cccc}
 & Exp. & Skyrme QRPA & LSSM \\
 \hline
%$\beta$        & 0.290 & 0.30 &  \\
B(E2)$\uparrow$$e^2b^2$ &0.1052(32)${\footnotemark[1]}$ & 0.11 &0.108\\
E [MeV]          & 3.628 & 2.71 & 3.54 \\
B(M1) $[\mu^2_N]$ & 1.113(49) & 1.52 & 0.826 \\
B(M1)$_{\rm{O}}$ $[\mu^2_N]$  & - & 0.60 & 0.316\\
B(M1)$_{\rm{S}}$ $[\mu^2_N]$  & - & 0.21 & 0.120\\
R                     & - & 2.9 & 2.6
\footnotetext[1]{From Ref.~\cite{bnl}.}
\end{tabular}
  \end{ruledtabular}
\end{table}

\subsubsection{Large-scale shell model}

The results of the LSSM calculations are presented in
Figure~\ref{shell}. We start with a reminder of the computational
procedures used in~\cite{sh1}, which we follow here. In the spherical
shell model, $^{50}$Cr is described in a 0$\hbar$$\omega$
space, i.e., ten particles are allowed to occupy all the states
available in the $fp$ shell. The KB3G~\cite{sh2} interaction is used.
The single-particle energies are taken from the
experimental spectrum of $^{41}$Ca.

Figure~\ref{shell} shows $B(M1)$ values obtained for the total (orbital
+ spin), orbital, and spin $M1$ transitions. In the spin part, a
quenching factor of 0.75~\cite{pvn} is used.  The calculations give
the state at 3.5~MeV with a sizeable strength $B(M1)=0.826\ \mu_N^2$.
The orbital strength dominates the 3.5~MeV state, which indicates
the scissors-like character of the state.
It should be pointed out that in the LSSM the 3.5-MeV state is not purely orbital but
also has some spin admixture, which appears in the QRPA calculations.
The calculated 3.5-MeV state matches remarkably well with the
experimentally observed state at 3628.2 keV.

The LSSM calculations show that  $M1$
transitions above 3.6 MeV are mainly of spin character (see lower
panel of Figure~\ref{shell}).
The calculated summed $M1$ strength in the experimentally investigated energy range from 2 - 9.7 MeV
amounts to $\sim$5.6 $\mu_N^2$ as compared to the measured value of 4.1(1) $\mu_N^2$.
The total calculated strength in the resonance, energy interval 4.5 - 14 MeV, is $\sim$6.6 $\mu_N^2$
in good correspondence to the QRPA result of 5.9 $\mu_N^2$.

%The calculated summed total B(M1) strength is $\sim$6.6 $\mu_N^2$ at the interval 4.5 - 14 MeV
%and $\sim$5.6 $\mu_N^2$  at 2 - 9.7 MeV (the energy range considered in the present experiment).
%These values are in an acceptable agreement with the Skyrme QRPA summed strengths 5.9 and 6.2 $\mu_N^2$,
%as well as with the experimental strength 4.01 $\mu_N^2$.

In Table~\ref{tab:Table3}, the LSSM results are compared to QRPA ones and
present experimental data.
The table shows that the LSSM better describes the level energy than
the Skyrme QRPA. Most probably this is because the LSSM takes into account
the coupling to complex configurations, omitted in the QRPA.

\section{Summary and outlook}

The low-lying $M1$-strength in the open-shell nucleus $^{50}$Cr
has been determined with the method of nuclear resonance fluorescence
using bremsstrahlung at the superconducting
Darmstadt linear electron accelerator S-DALINAC and
Compton backscattered photons at the High Intensity
$\gamma$-ray Source (HI$\gamma$S) facility.
Fifteen $1^{+}$ states have been observed between 3.6 and 9.7 MeV.
Further, 14 $\gamma$-ray transitions and 13 $B(M1)$ values have been measured for the first time.

The experimental results were compared to calculations in the
self-consistent Skyrme Quasiparticle Random-Phase-Approximation
(QRPA) \cite{Ring} and the Large-Scale Shell Model (LSSM) \cite{lssm_Gaurier}.
The QRPA allowed to highlight the basic features and origin of $1^{+}$ states
while the LSSM demonstrated the important role of complex configurations.
Both models reproduce the similar B(E2) values.
%nuclear deformation and gave
%qualitatively close results.

Following our theoretical analysis and comparison with other available
results for $fp$-shell nuclei the lowest $1^{+}$-state at 3.6 MeV was
identified as an isovector orbital mode with some spin admixture.
In agreement with previous studies for fp-shell nuclei~\cite{fe},
a constructive interference of the orbital and spin contributions for
this state was confirmed.
%It may thus be interpreted as the scissor mode in $^{50}$Cr.
This augments the previous rare data ($^{46,48}$Ti~\cite{ar} and $^{56}$Fe~\cite{fe})
on this orbital mode in the $fp$-shell region. The obtained data
match the estimations and trends for the scissors-like mode in a satisfactory way.

It is also interesting that distributions of
the orbital current, computed within the Skyrme QRPA
for the lowest $1^+$-state, favor the schematic picture of Lipparini and Stringari \cite{lip_str_83}
(isovector rotation-like oscillations inside the rigid surface)
rather than the scissors-like view of Lo Iudice and Palumbo \cite{Iudice}.

The spin excitations above the scissors-like mode were also inspected.
The QRPA calculation confirms that the spin $M1$ resonance in $^{50}$Cr is provided
by spin-flip transitions (mainly $1f_{7/2}-1f_{5/2}$) inside the $fp$-shell.

%{\bf Acknowledgments}
\begin{acknowledgments}
The effort of the S-DALINAC crew at TU Darmstadt
to provide an excellent electron beam is gratefully acknowledged.
We also thank the operators team of Duke Free Electron Laser Laboratory for providing excellent beams.
This work was supported by the DFG under grant No. SFB 634 and by the Helmholtz
International Center for FAIR. V.O.N. thanks the DFG RE 322/14-1, Heisenberg-Landau
(Germany-BLTP JINR), and Votruba-Blokhintsev (Czech Republic-BLTP JINR)
grants. J.K. and A.R. are grateful for the support
of the Czech Science Foundation (P203-13-07117S).
\end{acknowledgments}

\end{document}